\documentstyle[12pt]{article}
\def\ba{\begin{array}{c}}

\def\ea{\end{array}}

\def\bet{\beta}

\def\l{\left}
\def\l({\left(}
\def\r){\right)}
\def\r{\right}

\def\la{\lambda}
\def\al{\alpha}
\def\be{\begin{equation}}
\def\ee{\end{equation}}
\def\ed{\end{document}}
\def\bea{\begin{eqnarray}}
\def\eea{\end{eqnarray}}
\def\ll{\label}
\def\ni{\noindent}
\begin{document}
{\bf
\centerline 
{{\large Generation of Integrable  Quantum Nonultralocal Models}}
\centerline 
 {{\large through  Braided Yang-Baxter Equation}}
}
\centerline 
{Anjan Kundu}
\centerline 
{
Saha Institute of Nuclear Physics, Theory Group,}
\centerline 
{
 1/AF Bidhan Nagar, Calcutta 700 064, India}
\centerline 
{
 email: anjan@tnp.saha.ernet.in}
\begin{abstract}
Formulating quantum integrability  for  nonultralocal models (NM) parallel
to the familiar approach of inverse scattering
method is  a long standing problem. After reviewing our result regarding
 algebraic structures of ultralocal models, we look 
for
 the algebra underlying NM.
We propose an universal equation represented by braided  
 Yang-Baxter equation and
 able to  derive  all basic equations of
the known models like WZWN model, nonabelian Toda chain, 
quantum mapping  etc. 
 As further useful application
  we
  discover  new integrable  quantum
NM, e.g.   mKdV model, anyonic model,  Kundu-Eckhaus 
equation 
and derive SUSY models and reflection equation
 from the nonultralocal view point.
  
\end{abstract}
\smallskip

\ni 1.  {\bf Introduction }

Quantum integrable systems  (QIS)
 can be divided into two broad classes,
 namely {\it ultralocal} and {\it nonultralocal} depending on an important 
property 
of their 
representative Lax operators.
 Ultralocal QIS  are the 
standard and the most studied ones, which include well known models like
Nonlinear Schr\"odinger equation (NLS), sine-Gordon  model, Toda chain,
etc. They exhibit a common  ultralocality property that their Lax
operators at different lattice points $i \not =j$
commute: $[L_{1i}, L_{2j}]=0.$
  This fact is actively used 
in constructing their integrability theory expressed by the 
universal equation 
   \begin {equation}
R(\lambda , \mu)~ {L}_{1j}(\lambda)~ {L}_{2j}(\mu )
~ = ~  {L}_{2j}(\mu )~{L}_{1j}(\lambda)~ R(\lambda , \mu),
\ll{qybe}
\end {equation}
known as the quantum Yang Baxter equation (QYBE). 
Specific choices of  $L$ and  $R$ yield 
from (\ref{qybe}) the basic equations for concrete integrable models
\cite{fadrev}.
The underlying algebras  for  such
models also exhibit  common structure  related to 
 quantum algebras,
the knowledge of which
not only deepens our understanding of the system and adds beauty to the 
subject, but also helps to  generate
integrable 
nodels in a systematic way \cite{kb}.

However, it should be mentioned  that 
parallel development was not persuaded  
for the nonultralocal models (NM),
 characterised by the 
property $[L_{1i}, L_{2j}] \not = 0,$ though many famous models, e.g.
quantum KdV model, Supersymmetric models, nonlinear $\sigma$ models, WZWM
etc. belong to this class.
A    proposal in this line was made  in \cite{maillet}, 
 though  only for  models  convertable to ultralocal ones.
  A general formulation of the 
integrability theory given through 
a representative 
universal Yang-Baxter like equation 
is seriously lacking for such models
including the clear idea about the nature of the underlying algebraic 
structures. 
Our aim therefore is  to 
 look into  such systems from a rather general
point of view, where the  SUSY models,
 anyonic models as well as
models satisfying reflection type equation  
can also be considered as the representatives of the nonultralocal class
and  exploring the related algebra to 
 construct an extended
QYBE 
and generate concrete NM as particular cases,
parallel to the ultralocal models. 
\\ \\
2. {\bf Ultralocal models: underlying algebra 
 } 

We look first  into the established
  ultralocal
systems  for understanding  the role played by the underlying 
algebra.

Consider standard matrices $A$ and $B$ satisfying  the obvious
property $$A \otimes B= ( A_1  B_2 )= ( B_2 A_1)
 ~\mbox{or}  ~[A_1,B_2]=0~\mbox{ where}~ A_1 \equiv  A \otimes 1~\mbox{ and
}~ B_2= 
1 \otimes  B.$$ Let us choose now
 $A=L_i(\la),
B=L_i(\mu)$
 as Lax operators at the same lattice poins and check the above property.
It is immediate, that it no longer holds
due to the  operator nature of the
matrix elements of the quantum Lax operators  and the equation in effect 
turns
into the QYBE (\ref{qybe}), where a matrix  $R$ appears   to compensate
for the noncommutativity of matrix elements of $L$. This is the basic reason
for the appearance of 
nontrivial algebras underlying such integrable systems.

 On the
other hand  the choice   
 $A=L_{i+1}(\mu),
B=L_i(\la)$ does satisfy  the above commutation relation  due the
ultralocality property.

Let us now look into the standard matrix multiplication rule
   \begin {equation}
({A} \otimes  {B})(C \otimes D) = (AC \otimes BD)
\ll{mult}
\end {equation}
and check again for the Lax operators as 
 $A=L_{i+1}(\la),
 B=L_{i+1}(\mu),
 C=L_{i}(\la),
 D=L_{i}(\mu) . $ Note that since the multiplication rule (\ref{mult}) holds
due to the commutativity of $B_2$ and $C_1,$ that  also  remains valid
for the ultralocal Lax operators
yielding
\be
 (L_{1 i+1}(\la)
 L_{2 i+1}(\mu))
 (L_{1 i}(\la)
 L_{2 i}(\mu))
= (L_{1 i+1}(\la)
 L_{1 i}(\la))
( L_{2 i+1}(\mu)
 L_{2 i}(\mu))
\ll{mull}\ee
Now   coupling  two properties  (\ref{qybe})
and (\ref{mull}) of the local 
Lax operators one can derive a global  QYBE, essential for
proving the integrability which is a global property. Indeed,
starting from (\ref{qybe}) at $i+1$ point and multiplying with the same
relation at $i$ and subsequently using (\ref{mull}) one can globalise
the QYBE and repeating the step for $N$ times  obtain finally the
global equation
   \begin {equation}
R(\lambda , \mu)~ {T}_{1}(\lambda)~ {T}_{2}(\mu )
~ = ~  {T}_{2}(\mu )~{T}_{1}(\lambda)~ R(\lambda , \mu),
\ll{tqybe}\ee
for the monodromy matrix $T= \prod_i^N L_i.$ 
Trace $tr_{12}$ factorises the equation    yielding  $[tr T(\la), tr
T(\mu)]=0$ or the integrability condition 
$[C_n,C_m]=0,$ where $C_n, n=1,2,\ldots,N$ 
are the conserved quantities
generated by $\tau(\la)=Tr T(\la)$ as expansion coefficients.

To extract the algebra independent of the spectral parameters one can 
take  the ancestor Lax operator in the  form
\begin {equation}
L_{anc}^{q,t }( \xi) = L_{ij}, ~~~
 L_{11}=\xi {\tau_1^-} + \frac {1}{\xi}{\tau_1^+}, \ \ 
  L_{22}= \xi {\tau_2^-} +\frac {1}{\xi} {\tau_2^+}, \ \
 L_{12}=    {\tau_{21}} ,\ \
 L_{12}=
   {\tau_{12}} , 
\ll{anc}\end {equation}
along with  the twisted trigonometric
 $R(\la,\mu)$-matrix, which yields  the underlying 
  quadratic  algebra  
\begin {eqnarray}
t \  \tau_{12} \tau_{21} - t^{-1} \  \tau_{21} \tau_{12}  ~=~
- (q-q^{-1}) \left (  \tau_1^+ \tau_2^-  -  \tau_1^- \tau_2^+ \right )
 ~,  \nonumber \\
\tau_i^{\pm }\tau_{ij} ~=~q^{\pm 1 } t \ \tau_{ij} \tau_i^{\pm }~,~\ \
\tau_i^{\pm }\tau_{ji} ~=~q^{\mp 1 }  t^{-1} \  \tau_{ji} \tau_i^{\pm }~,~
\ll{etsa}\end {eqnarray}
for $i,j=(1,2) $. This Hopf algebra  exhibits a  coproduct structure
 and the multiplication rule (\ref{mult}), which are linked with the
transition  from  local to  global QYBE.
Taking different limits of $q,t$ and proper realisations of the algebra
(\ref{etsa}), from (\ref{anc}) 
one can generate different classes of  ultralocal models along with
 their exact Lax operators in a rather systematic way \cite{kb}.
\\ \\  3. {\bf Nonultralocal quantum systems: algebraic structure,
extended QYBE  and
 generation of  models}

We see immediately that both the above matrix  properties
fail for NM, since now the Lax operators   
do not  commute at the same as well as at different points. Therefore 
the  quantised algebra like (\ref{etsa})  with 
somewhat trivial multiplication property (\ref{mult}) needs
generalisation in the form 

   \begin {equation}
({A}\otimes  {B})(C \otimes D) =\psi_{BC} (A(C \otimes B)D)
\ll{multnm}
\end {equation}
where the noncommutativity of $B_2, C_1$ could be 
 taken into account. However the coproduct structure, essential
for the  globalisation of  QYBE must be preserved.
 Such idea realised in the  braided algebra \cite {majid} 
was implemented for formulating the integrability of NM \cite{kunhla}.
 Here we look into this problem 
from a bit different  angle and  report on new results.
Limiting ourselves to only two types of braidings (describing
 nearest neihgbours   by the matrix $Z$ and other neighbours
 by a single  braiding $\tilde Z$) we  present
 the braided generalisation of the
QYBE as

\begin{equation}
{R}_{12}(u-v)Z_{21}^{-1}(u,v)L_{1j}(u)\tilde Z_{21}(u,v)L_{2j}(v)
= Z_{12}^{-1}(v.u)L_{2j}(v) \tilde Z_{12}(v,u)L_{1j}(u){R}_{12}(u-v).
\ll{bqybel}\end{equation}
In addition this must be  complemented 
 by the  braiding relations  
\begin{equation}
 L_{2 j+1}(v)Z_{21}^{-1}(u,v)L_{1 j}(u)
=\tilde Z_{21}^{-1}(u,v)L_{1 j}(u)\tilde Z_{21}(u,v)
 L_{2 j+1}(v)\tilde Z_{21}^{-1}(u,v)
\ll{zlzl1u}\end{equation}
  at   nearest neighbour
points and
\begin{equation}
 L_{2 k}(v)\tilde Z_{21}^{-1}(u,v)L_{1 j}(u)
=\tilde Z_{21}^{-1}(u.v)L_{1 j}(u)\tilde Z_{21}(u,v)
 L_{2 k}(v)\tilde Z_{21}^{-1}(u,v)
\ll{zlzl2u}
\end{equation}
with $k>j+1$ 
  answering for the other neighbours.
Note that along with  the usual quantum $ R_{12}(u-v)$-matrix  
additional  $ \  \tilde Z_{12} , \ Z_{12}$ matrices 
 appear here, which can be (in-)dependent of the spectral parameters
and 
satisfy  a system of Yang-Baxter type relations
 \cite{kunhla}. 

The  set of relations (\ref{bqybel}-\ref{zlzl2u}) represent the universal 
equations for the integrable NM
within a certain class of braidings and 
particular choices for $R,L, Z, \tilde Z$  
derive concrete models of physical
interest. It is readily seen  that the trivial choice  $Z=\tilde
Z=1 $ reduces the above set into the standard QYBE (\ref{qybe}) together
with the ultralocality  condition, while  the nontrivial
 $Z$'s would lead to different types of NM.
 For example, 
the homogenious 
braiding
 $Z=\tilde Z$ would correspond to  
 SUSY,  anyonic models etc., while   
the choice $\tilde Z=1$ with  $Z \not = 1$ 
should describe the models like WZWN, quantum mKdV,
 nonlinear $\sigma$ models etc. with   $ \delta'$ 
function appearing in their fundamental commutation relations. The 
case $Z=1$ and nontrivial  $\tilde Z$ also appears to be consistent with 
the  reflection  
equation  \cite{sklya}.

For establishing integrability we have to construct first the QYBE for
the global monodromy matrix, which however can be carried out now almost
like the ultralocal case  due to  the changed multiplication rule
given by the braiding relations 
(\ref{zlzl1u})
and (\ref{zlzl2u}). For periodic models, where $L_1$ and $L_N$ become
 nearest neighbours some special care should be taken, which 
yields finally the global QYBE as 

\begin{equation}
{R}_{12}(u-v)Z_{21}^{-1}(u,v)T_{1}(u) Z_{12}^{-1}(v,u)T_{2}(v)
= Z_{12}^{-1}(v,u)T_{2}(v)  Z_{21}^{-1}(u,v)T_{1}(u){R}_{12}(u-v).
\ll{bgqybel}\end{equation}
Due to appearance of $Z$ matrices one faces initial difficulty in trace
factorisation, which nevertheless can be bypassed in most cases by
introducing a $K(u)$ matrix and    defining 
$t(u)=tr (K(u)T(u))$  \cite{sklya,kunhla}.

To demostrate the applicability of the present scheme we mention briefly 
the paricular explicit choices for the $R,Z,\tilde Z$ matrices, which
 derives 
 from the  general relations
  (\ref{bqybel}-\ref{zlzl2u}) the basic equations for the known models found
earlier and at the same time yield  new results. 
\\ 
 1. {\it Nonabelian Toda chain }\cite{natoda}
 
  $\tilde Z=1$  and $Z_{12}=
 {\bf 1} + i { h }(
 e_{22}\otimes e_{12})\otimes \pi$
and rational $R(u)$ matrix.\\
 2. {\it Current algebra in WZWN model} \cite{wzwn}
 
 $\tilde Z=1$ and $Z_{12}=R_{q12}^-$ 
\\
 3. {\it Coulomb gas picture of  CFT} \cite{babelon}
 
  $\tilde Z=1$ and
 $Z_{12}=
q^{-\sum_i H_i
\otimes H_i}
$.
\\
 4. {\it Nonultralocal quantum mapping }\cite{Nijhof}

 $\tilde Z=1$ and $Z_{12}(u_2)
 = {\bf 1}+ \frac { h }{u_2}\sum_\alpha^{N-1}e_{N \alpha
}\otimes e_{\alpha N}~~~$
and rational $R(u)$.
 \\5. {\it Integrable model on moduli space }\cite{alex}

 Homogenious braiding with $\tilde Z=Z_{12}=R^+_q$ and trigonometric
$R(u)$ matrix.
\\ \\ 
 { \bf 4.  New results}
 \\
1. {\it Supersymmetric models } \cite{SUSY}

We shall look into the SUSY models from the NM point
and  derive their basic equations from our formulas by choosing 
 $Z=\tilde Z=\sum \eta_{\al \bet} g_{\al \bet}$, where $ \eta_{\al \bet}= 
e_{\alpha \alpha}\otimes e_{\beta \beta}$ and $ g=
 (-1)^{\hat \alpha \hat
 \beta}$ with  supersymmetric grading  $\hat \alpha.$
 $R$-matrix is either  rational
or trigonometric of $SU(1,1)$ type.
\\
2. {\it Reflection equation}

The reflection equation of \cite{sklya} can also be derived  from the
nonultralocal point of view by considering a monodromy matrix $T=\prod_j
\tilde L_j$ such that 
$$\tilde L_j(u)= L_{j-N}(u),\mbox{for} ~ j >N,~~ = K^-(u), \mbox{for} 
~j=N ,~~\mbox{and}~~ = L^{-1}_{N-j}, \mbox{for}~  j<N.$$
 Though this refers to a bit different
braiding than that we considered here, the choice $Z=1, \tilde
Z=R(u_1+u_2)$ and  without periodicity recovers most of the results.
\\
3. {\it Anyonic SUSY model}

Generalizing the SUSY model  we choose
$Z=\tilde Z=  
 \sum \eta_{\al \bet} \tilde g_{\al \bet}$, where  
  an anyonic phase is included in $\tilde g_{\al \bet}=e^{i \theta 
\hat \alpha \hat
 \beta}$. Rational  $R$ matrix derives integrable anyonic moels while 
the trigonometric one  can describe  the $q$-deformed
anyons.
\\ 4. {\it Kundu-Eckhaus equation} \cite {kun84}

This classically integrable  NLS equation with 5th power nonlinearity  
\be i\psi_1+\psi_{xx}+ \kappa (\psi^\dagger \psi) \psi
+ \theta^2   (\psi^\dagger \psi)^2 \psi +2i \theta
 (\psi^\dagger \psi)_x \psi =0, \ee

as a  quantum model involves fields with a  nice anyonic type  
representation: $$ ~\psi_n \psi_m= e^{i \theta} \psi_m\psi_n,~~ n>m;  
~~[ \psi_m, \psi_n^\dagger]=1.$$
Our scheme gives the solution $\tilde Z=1,
Z= diag (e^{i\theta},1,1,e^{i\theta})$ and the rational $R$ matrix.
Though we are able to  construct the braided QYBE,
 the trace factorisation problem
 could not be solved due to some pecularity of the
monodromy matrix. 
\\
\ni 5. {\it Quantum mKdV model  }
 
An exact quantum solution of this interesting model can be found 
 following  the present  scheme  with 
 $\tilde Z=1, ~
Z_{12}=  Z_{21}= q^{-\frac {1}{2} \sigma^3\otimes \sigma^3
},$
and  the  trigonometric $R(u)$ matrix. For details see \cite{kmpl96}, while 
the details of the other results will be given elsewhere.  
\smallskip

The author expresses his sincere thanks to the Alexander von  Humboldt
 Foundation for the ressumption of  fellowship
 (in the Phys. Inst., Bonn
University) and the financial support.

\ed
\end{document}